\pdfoutput=1
\documentclass[11pt]{article}
\usepackage[final]{acl}
\usepackage{times}
\usepackage{latexsym}
\usepackage[T1]{fontenc}
\usepackage[utf8]{inputenc}
\usepackage{microtype}
\usepackage{amsmath}
\usepackage{braket}
\usepackage{subfigure}
\usepackage[export]{adjustbox}
\usepackage{url}
\usepackage{bm}
\usepackage{multirow}
\usepackage{tikz}
\usepackage[ruled,vlined,algo2e]{algorithm2e}
\usepackage{algorithm}
\usepackage{algorithmic}
\usepackage{graphicx}
\usepackage{setspace}
\usepackage{hyperref}
\usepackage{enumitem}
\usepackage{makecell}
\let\Algorithm\algorithm
\renewcommand\algorithm[1][]{\Algorithm[#1]\setstretch{1}}

\usepackage{xcolor}
\usepackage[ruled,vlined,algo2e]{algorithm2e}

\SetCommentSty{mycommfont}
\AtBeginDocument{
  \let\mathbb\relax
  \DeclareMathAlphabet{\mathbb}{U}{msb}{m}{n}
}

\usepackage{amsmath}
\usepackage{booktabs}
\usepackage{amssymb}% http://ctan.org/pkg/amssymb
\usepackage{pifont}% http://ctan.org/pkg/pifont
\newcommand{\cmark}{\ding{51}}%
\newcommand{\xmark}{\ding{55}}%

\usepackage{multirow}
\usepackage{multicol}
\usepackage{adjustbox}
\usepackage{arydshln}
\usepackage{tcolorbox}

\definecolor{myblue}{HTML}{4E84C4}
\definecolor{myred}{HTML}{B02418}
\definecolor{mygreen}{HTML}{34692E}
\definecolor{myorange}{HTML}{DA7842}
\definecolor{paperblue}{HTML}{077dea}
\definecolor{babyblue}{HTML}{E3EDF7}

\newcommand{\coloredalpha}{\textcolor{paperblue}{\alpha}}
\newcommand{\coloredgamma}{\textcolor{paperblue}{\gamma}}
\newcommand{\coloredsigma}{\textcolor{paperblue}{\sigma}}
\newcommand{\coloreddelta}{\textcolor{paperblue}{\delta}}

\makeatletter
\patchcmd{\maketitle}
 {\def\@makefnmark}
 {\def\@makefnmark{}\def\useless@macro}
 {}{}
\makeatother

\title{AppBench: Planning of Multiple APIs from Various APPs \\ for Complex User Instruction}

\author{Hongru Wang$^{\coloredalpha\dagger}$\thanks{$^\dagger$ Equal Contributions. Work done when first author visits EdinburghNLP.}, Rui Wang$^{\coloredalpha\dagger}$, Boyang Xue$^{\coloredalpha}$, Heming Xia$^{\coloredgamma}$, \\ \bf Jingtao Cao$^{\coloredalpha}$, Zeming Liu$^{\coloredsigma}$, Jeff Z. Pan$^{\coloreddelta\ddagger}$\thanks{$^\ddagger$ Co-corresponding Authors}, Kam-Fai Wong$^{\coloredalpha\ddagger}$ \\
  $^{\coloredalpha}$The Chinese University of Hong Kong \\
  $^{\coloredgamma}$The Hong Kong Polytechnic University \\
  $^{\coloredsigma}$Beihang University,
  $^{\coloreddelta}$The University of Edinburgh  \\
  \texttt{hrwang, kfwong@se.cuhk.edu.hk} }
\date{}

\begin{document}
\maketitle
\begin{abstract}
Large Language Models (LLMs) can interact with the real world by connecting with versatile external APIs, resulting in better problem-solving and task automation capabilities. Previous research primarily focuses on APIs with limited arguments from a single source or overlooks the complex dependency relationship between different APIs. However, it is essential to utilize multiple APIs collaboratively from various sources (e.g., different Apps in the iPhone), especially for complex user instructions. In this paper, we introduce \texttt{AppBench}, the first benchmark to evaluate LLMs' ability to plan and execute multiple APIs from various sources in order to complete the user's task. Specifically, we consider two significant challenges in multiple APIs: \textit{1) graph structures:} some APIs can be executed independently while others need to be executed one by one, resulting in graph-like execution order; and \textit{2) permission constraints:} which source is authorized to execute the API call. We have experimental results on 9 distinct LLMs; e.g., GPT-4o achieves only a 2.0\% success rate at the most complex instruction, revealing that the existing state-of-the-art LLMs still cannot perform well in this situation even with the help of in-context learning and finetuning. Our code and data are publicly available at \url{https://github.com/ruleGreen/AppBench}.

\end{abstract}

% it is necessary to utilize multiple APIs collaboratively from various sources to complete the complex user instructions, call for demand to consider the \textit{graph structures} (i.e., some APIs can be executed independently while others need to be executed one by one) and \textit{permission constraints} (i.e., the authorized source to execute) issues. To address these issues, we introduce \texttt{AppBench}, the first benchmark to evaluate LLMs' ability to plan and execute multiple APIs from various sources in order to automatically complete user's complex instructions, considering four different levels of planning: Single APP Single API (SS), Single APP Multiple API (SM), Multiple APPs Single API (MS), Multiple APPs Multiple API (MM).

\section{Introduction}
Empowering Large Language Models (LLMs) \cite{zhao2023survey} with versatile tools such as retrievers \cite{wang-etal-2023-large, wang-etal-2024-uniretriever}, models \cite{shen2023hugginggpt}, and even physical robots \cite{liang2023code}, holds significant promise in overcoming inherent limitations, such as hallucination \cite{ji2023survey} and outdated information \cite{nakano2021webgpt, liu2023webglm}, and unveils the immense potential for LLMs to tackle increasingly complex and interactive real-world tasks \cite{li-etal-2023-api, lu2023chameleon}. Over the past several months, lots of new benchmarks and datasets have been proposed to evaluate the performance of different LLMs to adeptly select and execute various tools \cite{li-etal-2023-api, shen2023hugginggpt, huang2024planning}, marking a pivotal milestone in their evolution. Out of plentiful tools in practice, APIs have become one of the fundamental and promising tools in today's digital world, due to greater flexibility and customizability with well-defined format and ease of execution \cite{qin2023tool}. 

\begin{figure}[t]
    \centering
    \includegraphics[trim={11cm 5.5cm 9.5cm 4cm}, clip, width=0.5\textwidth]{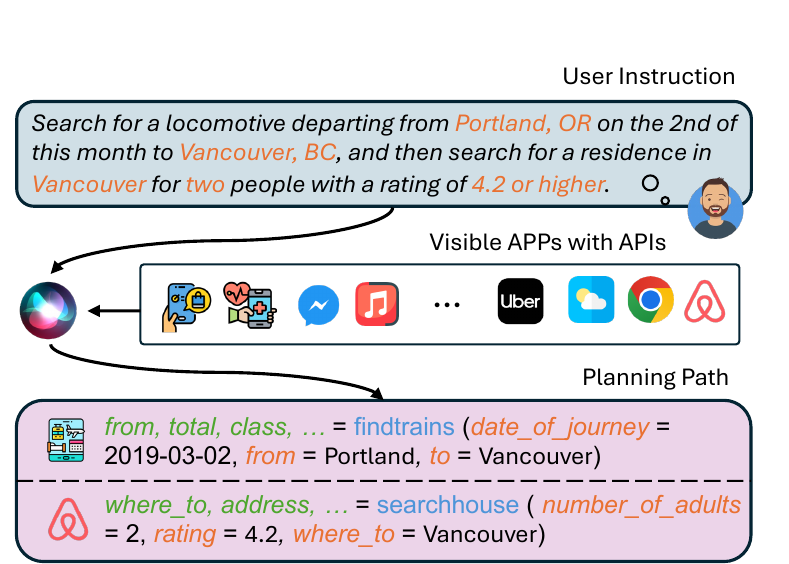}
    \captionsetup{font=10pt}
    \caption{An example of one user instruction requires two independent APIs from different APPs since input arguments of both two APIs do not rely on each other. We use different icons to indicate different APPs, and color \textcolor{myblue}{API}, and \textcolor{mygreen}{returned arguments} and \textcolor{myorange}{input arguments}.}
    \label{intro_example}
    \vspace{-7mm}
\end{figure}

% two points: 1. graph structure; 2. data isolation
Previous works have attempted to evaluate LLMs on their ability to call the correct API in multiple turn dialogues, such as API-Bank \cite{li-etal-2023-api} and ToolBench \cite{qin2023toolllm}, or single turn instructions, like APIBench \cite{patil2023gorilla}. However, most existing benchmarks focus either on a single API call in a single turn or on APIs with limited arguments. For instance, API-Bank mainly evaluate one API call per turn in multi-turn dialogues, while APIBench and ToolBench considers APIs only with one or two arguments (e.g., only one output with one or two inputs). Furthermore, the small number of arguments makes it difficult to fully explore the complex dependency relationships between multiple APIs. For instance, the input arguments for a current API may depend on the return arguments of several previous APIs. These limitations highlight a gap in addressing complex user instructions when it is necessary to utilize multiple APIs in practice, underscoring the need for more comprehensive and practical evaluation benchmarks. 

To bridge the gap, we introduce a new evaluation benchmark: \texttt{AppBench}, representing the first effort to assess the aptitude of LLMs to function as the meta planner for multiple APIs from various sources for complex user instruction. Specifically, we simulate a situation in which the user instruction can be fulfilled through collaboratively API calls from various APPs in the mobile device. Figure~\ref{intro_example} shows one typical example. Given the complex user instruction, the meta LLM, such as Apple's Siri and Google Assistant, need to plan an executable path according to user instruction and corresponding API descriptions. To fulfill this requirement, it is necessary not only to indicate which APP will distribute and execute each API but also to specify the execution order of the APIs, including all necessary inputs and returned arguments. We consider this setting aligns well with the complexity and practical limitations in the real world, and presents a great opportunity for advanced AI assistants like Apple's Siri to showcase their intelligence and capability in orchestrating collaborative API executions across multiple Apps.

In this way, two significant challenges are identified: \textit{graph structure} and \textit{permission isolation}. Firstly, the inter-dependency between multiple APIs creates a more complex execution structure. Some APIs can be executed independently, while others are dependent and must be executed sequentially, resulting in a graph-like structure. Secondly, these APIs may originate from different sources, and the LLM might not have permission to call them directly. This necessitates identifying the authorized source for each API. For instance, APIs from one company may only be executed by an LLM within the same company.  In doing so, we aim to chart a path towards realizing the vision of an intelligent assistant capable of seamlessly navigating and interfacing with the myriad APPs and APIs pervasive in contemporary digital ecosystems. To conclude, our contribution can be summarized in three folds:

\begin{itemize}
    \item To the best of our knowledge, we are the first to identify graph structure and permission isolation issues of multiple API calls when addressing complex user instructions.

    \item We propose \texttt{AppBench}, serving as an important complementary evaluation benchmark to assess the planning capabilities of different LLMs as meta planner for these APIs. Additionally, we introduce an automatic data collection pipeline, which can be used to gather data efficiently and effectively.
    
    \item Our experimental results on 9 distinct LLMs demonstrate almost all models, including the latest GPT-4o, fall short in this setting, particularly when dealing with complex graph planning structures. Further analysis shows that simple in-context learning and fine-tuning do not significantly improve performance.
\end{itemize}

\section{Related Work}

\paragraph{Tool Benchmarks.} The complexity of real-world tasks necessitates the integration of diverse tools and services, consisting of three types of tools \cite{qin2023tool}: 1) physical interaction-based tools \cite{liang2023code}; 2) GUI-based tools \cite{wang2024mobileagent}; and 3)
program-based tools \cite{wang-etal-2023-large, li-etal-2023-api}. On the one hand, some work focuses on models, retrievers, or calculators to address the intrinsic limitations of LLMs, such as ToolQA \cite{zhuang2023toolqa} and ToolBench \cite{qin2023toolllm}. 
On the other hand, another line of work targets APIs since they are particularly crucial for bridging smooth interaction between humans and the digital realm \cite{li-etal-2023-api, qin2023toolllm, huang2024planning}. 
Most previous works formulate this as an API selection task given all related information about each API and current input, which overlooks the nuanced dependencies and permission constraints between different APIs, such as APIBench \cite{patil2023gorilla} and API-Bank \citep{li-etal-2023-api}. Nevertheless, the successful execution of APIs in the real world necessitates meeting requirements fulfilled (either the value is provided by the user or previous APIs) and obtaining permission from trusted agents beyond just knowing API names and a few arguments. More details can refer to latest survey \cite{qu2024toollearninglargelanguage} and tutorial \cite{tool_tutorial}.

% Our framework addresses this complexity by enabling language agents to intelligently interact with APIs across applications. Incorporating permission isolation mechanisms and understanding API dependencies, our approach ensures robust tool learning, maximizing language agent potential in real-world contexts.

\paragraph{Language Agent.} Existing frameworks for language agents have made notable strides in facilitating interaction with external tools \cite{shen2023hugginggpt, li-etal-2023-api, huang2024planning} and environment \cite{VirtualHome, wang-etal-2022-scienceworld}. They usually follow the single-agent paradigm to access different tools or services sequentially \cite{lu2023chameleon, li-etal-2023-api}, or multi-agent framework by assigning different agents different roles to call different cognitive tools \cite{tpe} . For example, \citet{lu2023chameleon} propose Chameleon which utilizes one agent to plan the execution order of different services by outputs a sequence of names of tools, which assume that the agents to call these tools are already known, and lots of works follow \cite{xu2023rewoo, huang2024planning}. Furthermore, various benchmarks are proposed to evaluate the abilities of LLMs serving as agents in different situations \cite{li-etal-2023-api, liu2023agentbench, ma2024agentboard}. For instance, \citet{yao2023webshop} proposes WebShop to evaluate whether LLMs are capable of interacting with the Web. Similarly, \cite{puig2018virtualhome} simulates household activities through programs, and many works use this as a testbed for embodied agents \cite{hao2024toolkengpt}. Latest work focus on using APIs or functions to control the whole planning processing of agents \cite{selfdc}.

\section{\texttt{AppBench} Construction}

\begin{figure}[t]
    \centering
    \includegraphics[trim={5cm 6cm 15cm 4cm}, clip, width=0.85\textwidth]{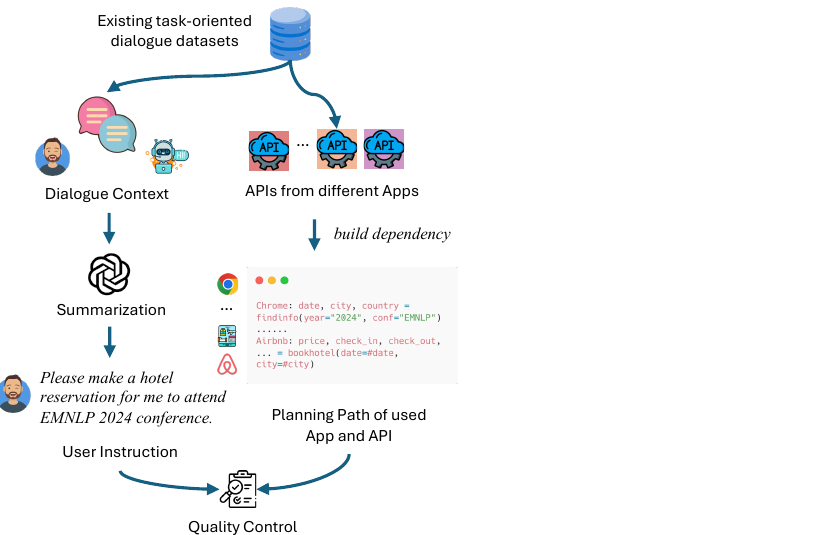}
    \captionsetup{font=10pt}
    \caption{A high-level processing to collect the \texttt{AppBench}, taking advantages of existing task-oriented dialogue datasets.}
    \label{data_collection}
\end{figure}

In this section, we start with a formal task definition and then provide a detailed explanation of how we efficiently and effectively built our \texttt{AppBench} by leveraging existing datasets.

\subsection{Task Definition}

Given the user instruction $u$ and a virtual mobile environment with an APP family, $\mathcal{E} = \{APP_1, APP_2, ..., APP_n\}$ where each APP contains several APIs $\{p_i^1, .. p_i^j\}$ where $i$ stands for $i_{th}$ APP and $j$ means $j_{th}$ API inside this APP, the meta agent need to decide an executable path to call different APIs from various APPs to fulfill the instruction in the format of the list which each item in the list is $\{APP_i: r_1, r_2, .., r_m = p_i^j (k_1=v_1, ..., k_n=v_n)\}$. 
The $APP_i$ and $p_i^j$ denote the name of the APP and corresponding API of this APP, and the $r_i$ and $k_i$ mean the $i_{th}$ returned and input arguments respectively. The $v_i$ can be the actual value provided by the user or a returned argument by previous APIs.

\subsection{Data Categorization}

Based on the number of APPs and APIs utilized in each user instruction, the data can be categorized into four distinct types. Each category represents a typical use case in practical scenarios, creating a comprehensive benchmark for evaluating real-world applications when combined.

\begin{itemize}[leftmargin=*]
    \item \textbf{\textit{Single APP Single API (SS)}} The instructions of the users only need to utilize one API from one APP.

    \item \textbf{\textit{Single APP Multiple API (SM)}} The instructions of the users need to utilize multiple API from one APP. It is important to note that these APIs can be called either sequentially or concurrently, depending on whether there is a dependency between their arguments.

    \item \textbf{\textit{Multiple APPs Single API (MS)}} The instructions of the users need to utilize multiple APIs and each of them belongs to one different APP. Also, there may exist dependency between APIs across different APPs.

    \item \textbf{\textit{Multiple APPs Multiple API (MM)}} The instructions of the users need to utilize multiple APIs and multiple APPs. The difference with MS is there may exist multiple APIs come from the same APP. Furthermore, the dependency relationship between arguments can be the most complex when dealing with APIs from the same APP or from different APPs.

\end{itemize}

Table~\ref{tab:existing_work} shows the detailed comparison between \texttt{AppBench} with other popular benchmarks. Most of existing benchmark focus on part of these typical situations or overlook the complex dependency relationships between multiple APIs. In addition, our formulation highlights the potential for investigating \textit{graph structure} and \textit{permission management}, considering the inherent complexity of APIs and Apps, particularly in terms of handling DP in multiple input and output arguments.

\begin{table}[!t]
\setlength{\belowcaptionskip}{0pt}
    \centering
    \begin{adjustbox}{max width=0.45 \textwidth}
    \begin{tabular}{l|ccccc}
    \toprule
    \textbf{Benchmark} & \textbf{SS} & \textbf{SM} & \textbf{MS} & \textbf{MM} &\textbf{DP} \\
    \hline
    APIBench \citep{patil2023gorilla} &  \cmark & \xmark & \xmark & \xmark & \xmark \\
    API-Bank \citep{li-etal-2023-api} & \cmark & \xmark & \xmark & \xmark& \xmark \\
    ToolQA \citep{zhuang2023toolqa} & \cmark &  \xmark & \xmark & \xmark & \xmark \\
    ToolBench \citep{qin2023toolllm} & \cmark & \cmark & \cmark & \cmark & \xmark \\
    UltraTool \citep{huang2024planning} &\cmark & \cmark & \xmark & \xmark & \cmark \\
    \hline
    AppBench (Ours) & \cmark & \cmark & \cmark & \cmark & \cmark \\
    \bottomrule
    \end{tabular}
    \end{adjustbox}
    \caption{Comparison with existing evaluation benchmarks at the turn-level for a fair comparison. DP stands for Dependency.}
    \label{tab:existing_work}
    \vspace{-6mm}
\end{table}

% trim={1cm 2cm 0cm 0cm}, clip,
\begin{figure*}[t]
    \centering
    \includegraphics[trim={1cm 2cm 0cm 0cm}, clip, width=0.9\textwidth]{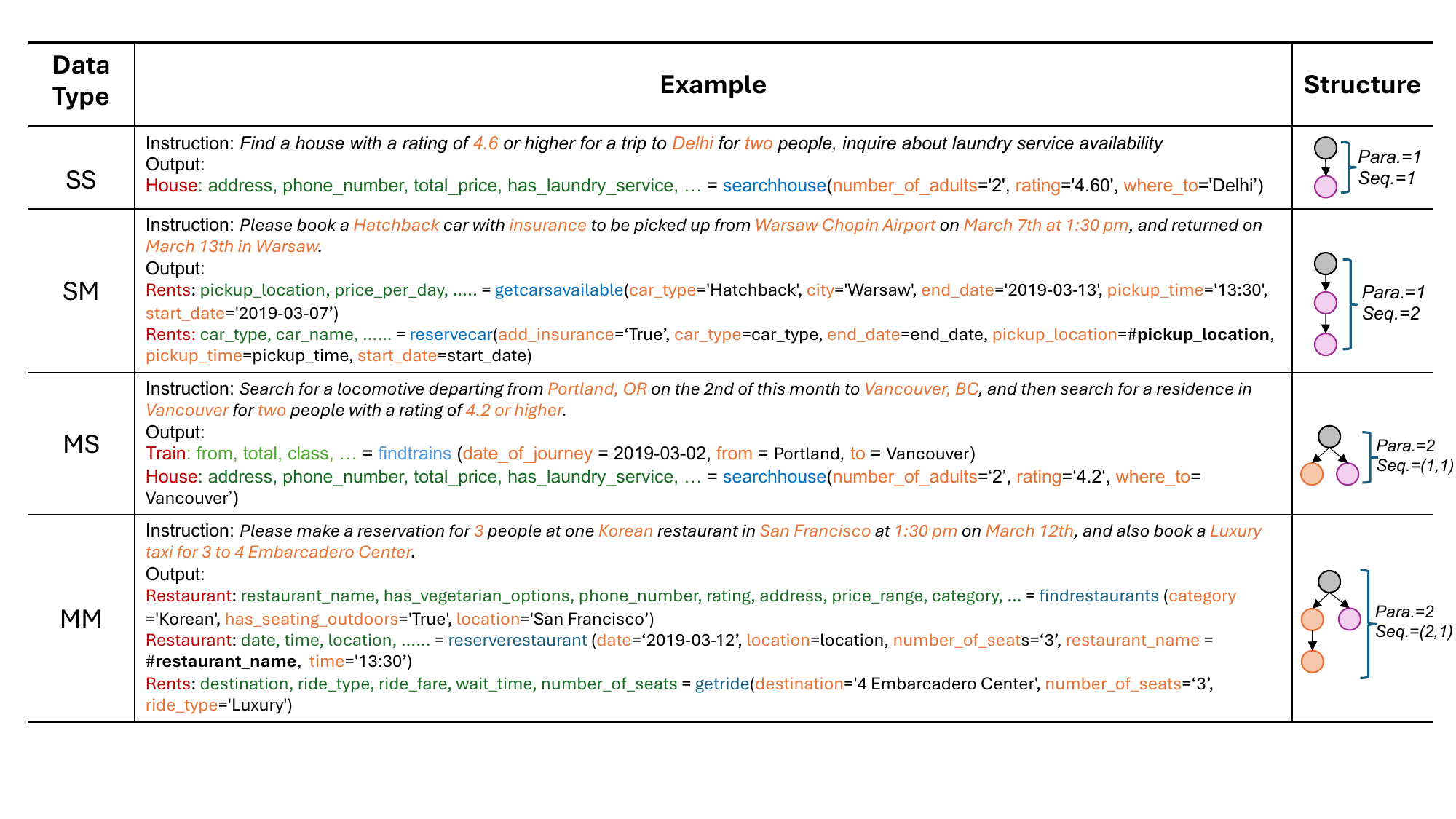}
    \captionsetup{font=10pt}
    \caption{An example of different types of samples in \texttt{AppBench}. We color \textcolor{myred}{APP}, \textcolor{myblue}{API}, and \textcolor{mygreen}{returned arguments} and \textcolor{myorange}{input arguments}. We also present the structure of the example using grey nodes and colorful nodes to indicate user instruction and APIs from different APPs, respectively. We bold the \textbf{argument} which is returned by the previous API call (a.k.a., dependency relationship). Para. and Seq. represents the parallel and sequential size of the corresponding data sample. We emphasize we only choose the simplest examples in each type for better understanding, there are data samples with much more complex logic structures in the original dataset.}
    \label{data_examples}
\end{figure*}

\subsection{Data Collection}
To maximize the authenticity of user instructions and minimize human efforts, we prioritize using existing task-oriented dialogue datasets \cite{rastogi2020towards, budzianowski2018multiwoz}. These datasets are typically collected through human-to-human interactions in real-world scenarios and contain a wide range of APIs across numerous domains and services. Specifically, we selected the SGD \cite{rastogi2020towards} dataset as the seed dataset because it encompasses most of the domains and APIs. We then utilized LLMs and Python scripts to generate the desired inputs and outputs, respectively. Figure~\ref{data_collection} illustrates the detailed procedures.

\paragraph{Instruction Acquisition.} Firstly, we extract the utterances of the user and system in the task-oriented dialogue and feed it into the LLM\footnote{GPT-4o during the collection} to summarize the user's requirements in one instruction. 
For example, the user may want to know the city and date of EMNLP 2024, and book a hotel according to the city and date. 
In the previous task-oriented dialogue, this is achieved by multi-turn interactions. 
In contrast, we summarize the whole dialogue into one complex user instruction to mimic more natural and complex cases in practice. To ensure that the values of certain intermediate arguments (such as \textit{date} and \textit{city}) are not disclosed at the instruction, we require the LLM to avoid outputting the actual values of other arguments, except for those that are explicitly provided in the prompts, such as user-aware arguments. The prompt details can be found in Appendix~\ref{appendix:data_collection} \footnote{All prompts can be found in Appendix if not stated.}.

\begin{table}[!t]
\setlength{\belowcaptionskip}{0pt}
    \centering
    \begin{adjustbox}{max width=0.9\textwidth}
    \begin{tabular}{l | cccc}
    \toprule
    % dynamic / multi-turn?
    \textbf{Statistics} & \textbf{SS} & \textbf{SM} & \textbf{MS} & \textbf{MM} \\
    \hline
    \# Samples & 200 & 200 & 200 & 200 \\
    \# Apps & 9 & 11 & 10 & 11 \\
    \# APIs & 11 & 22 &12 & 23\\
    \hline
    % Max Apps & 1 & 1 & 4 & 4\\
    % Max APIs & 1 & 4 & 4 & 8\\
    % Max. arguments & 6 & 7 & 6 & 7\\
    Avg. Apps & 1.0 & 1.0 & 2.7 & 2.2 \\
    Avg. APIs & 1.0 & 2.2 & 2.7 & 3.3 \\
    Avg. arguments & 4.0 & 4.5 & 3.8 & 4.4 \\
    \hline
    
    Max. Seq. & 1 &4 &4& 8 \\
    Max. Para. & 1 &1 &4 & 3\\
    
    Avg. Seq. & 1 &2.2 &1.2& 1.9 \\
    Avg. Para. & 1 &1.0 &2.2 & 1.8\\
    
    \bottomrule
    \end{tabular}
    \end{adjustbox}
    \caption{The data statistics of our proposed \texttt{AppBench}.}
    \label{tab:data_stat}
    \vspace{-4mm}
\end{table}

\paragraph{Planning Path.} Besides the instruction part, we write a Python script to automatically parse the API calls at different system turns in the multi-turn dialogue to form the planning path as the output. Specifically, we regard different domains (a.k.a., services) in task-oriented dialogue as different APPs such as restaurants and hotels, and extract the name of the domain and API first to locate which APP should invoke to call the API, and then we follow the execution order of different APIs to build the dependency between various arguments. For example, if the returned arguments from the previous API are required in the current API, we use \#name to indicate it such as \#date and \#city in the Figure~\ref{data_collection}. In this way, we can get an executable and unique path to execute APIs from different APPs.

\paragraph{Quality Assessment} To ensure the quality of data, we utilize a Python script to validate whether or not all actual values are provided from the user side, and none of values are provided from the system side. Furthermore, we adapt GPT-4o to score each instruction in terms of fluency and diversity from 1 to 10, and then remove cases whose score is lower than 6. Approximately 20\% of the samples were removed, and the average score of the remaining samples is around 8.05. We finally manually check each instruction-path pair, and remove some mismatch pairs such as the instruction is simple or API calls can not complete the user instructions, resulting in 200 high-quality samples for each category. 

% Since the output is built automatically and grounded on labels in the original dialogue dataset, it can be assumed to be correct naturally. Alternatively, to

\subsection{Data Statistic}

Table~\ref{tab:data_stat} illustrates the statistics of \texttt{AppBench}. Specifically, there are approximately 10 different APPs for each type and over 20 various APIs in both MS and MM. We provide the list of all APP and API in Table~\ref{tab:apps_apis}. Secondly, the average number of APIs increases from SS, SM to MS, MM, revealing the complex relationship. We also emphasize that the higher number of arguments for each API aligns with the complicated nature of tool execution in practice, as there may be multiple input and returned arguments for one API. Furthermore, we provide statistics about sequential and parallel relationships in each category (Seq. and Para.), revealing the complex graph structure in the dataset. Figure~\ref{data_examples} presents one example for each category for better understanding. More analysis can be found in the Appendix~\ref{appendix_data_stat}.

% We don't consider the response 1) the middle planning is more important, if it is wrong, then the response becomes useless; 2) given the correct planning path to call APIs, the response can be easily generated for current LLMs.

% And 

% These criteria offer a comprehensive understanding of the structural dynamics at play, shedding light on the intricate web of API dependencies and their implications for task execution.

% For example, a dependency from API-A to API-B instigates a directed edge running from API-A to API-B. 

% To quantify the complexity inherent in these interactions, we compute both the average and maximum \textbf{sizes} of these components (\textit{Max. and Avg. Compo.} in Table~\ref{tab:data_stat}), which reflect the complexity of dependency among the APIs. 
% Additionally, we measure both the average and the maximum \textbf{number} of components within each sample (\textit{Max. and Avg. \# Compo.} in Table~\ref{tab:data_stat}), providing insight into the level of parallelism among the APIs or components. 
% These criteria offer a comprehensive understanding of the structural dynamics at play, shedding light on the intricate web of API dependencies and their implications for task execution.

\section{Experiments}

\subsection{Setup}

\paragraph{Models.} We choose several LLMs from both open- and closed-source models, aiming to provide a comprehensive evaluation, following \cite{huang2024planning, zhuang2023toolqa}. Specifically, we choose Mistral-7B (\texttt{Mistral-7B-v0.2}) \cite{jiang2023mistral}, the LLaMa3 series \citep{llama3modelcard} (\texttt{Meta-Llama-3-8B/70B-Instruct}), and the Qwen series \citep{qwen} (\texttt{Qwen1.5-7B/14B/72B-Chat}) from open-source LLMs.  Besides that, we also select GPT3.5 (\texttt{gpt-3.5-turbo}) and GPT4 (\texttt{gpt-4o}) from closed-source LLMs. We also tried other models such as LLaMA2-7B or Vicuna but we find it difficult for them to output in the required format.

\paragraph{Implementation Details.} We set the temperature and top p as 0.1 to reduce randomness. The experiments of open-source models are run on NVIDIA A100 GPUs and those of closed-source models are fulfilled by APIs of OpenAI. To address the limitations imposed by the varying context windows of different LLMs, we adopt a \textit{hierarchical} prompting approach. First, we prompt the LLMs to identify the relevant APP. Once the appropriate APP is determined, we then provide the LLMs with only the API descriptions of these specific APPs.

\begin{table*}[!t]
\setlength{\belowcaptionskip}{0pt}
    \centering
    \begin{adjustbox}{max width=0.96 \textwidth}
    \begin{tabular}{l| ccc|ccc|ccc|ccc}
    \toprule
    \multirow{2}{*}{\textbf{Models}} & \multicolumn{3}{c}{\textbf{SS}}  & \multicolumn{3}{|c}{\textbf{SM}} & \multicolumn{3}{|c}{\textbf{MS}}  & \multicolumn{3}{|c}{\textbf{MM}} \\    
    % dynamic / multi-turn?
    \cline{2-13} & $F1_{app}$ & $F1_{api}$ & Succ & $F1_{app}$ & $F1_{api}$ & Succ & $F1_{app}$ & $F1_{api}$ & Succ & $F1_{app}$ & $F1_{api}$ & Succ \\
    \hline
    Mistral-7B & 55.97 & 16.31 & 0.51 & 36.59 & 15.09 & 0.50 & 33.72 & 6.42 & 0.00 & 28.92 & 7.56 & 0.00 \\
    Vicuna-13B & 43.20 & 3.70 & 2.00 & 34.71 & 4.63 & 0.50 & 20.43 & 3.10 & 0.00 & 21.05 & 2.52 & 0.00 \\
    \hdashline
    
    LLaMA3-8B & 63.04 & 42.67 & 23.23 & 37.20 & 25.33 & 0.50 & 30.65 & 19.52 & 0.10 & 26.39 & 17.80 & 0.05  \\
    LLaMA3-70B & 71.20 & \underline{70.00} & \underline{50.00} & 46.48 & \underline{46.96} & \underline{10.50} & 32.61 & 32.96 & 2.50 & 28.97 & \underline{28.53} & 0.50 \\
    
    \hdashline
    QWen1.5-7B & 48.14 & 19.54 & 0.00 & 30.13 & 16.71 & 0.00 & 23.24 & 10.11 & 0.00 & 23.76 & 11.55 & 0.00 \\
    
    QWen1.5-14B & 72.89 & 28.41 & 10.10 & 41.89 & 25.51 & 1.50 & \underline{42.22} & 21.98 & 0.80 & 32.36 & 15.07 & 0.00 \\
    QWen1.5-72B & \underline{81.23} & 24.28 & 12.50 & \textbf{51.89} & 25.27 & 1.00 & \textbf{45.94} & 13.42 & 0.62 & \textbf{38.53} & 11.51 & 0.00 \\

    \hdashline
    
    GPT-3.5 & 63.60 & 57.95 & 30.81 & 41.49 & 43.65 & 6.50 & 33.17 & \underline{34.53} & \underline{7.00} & 27.79 & 28.09 & \underline{1.00} \\
    GPT-4o & \textbf{88.31} & \textbf{86.87} & \textbf{70.92} & \underline{50.83} & \textbf{50.57} & \textbf{20.50} & 39.39 & \textbf{39.14} & \textbf{11.00} & \underline{32.62} & \textbf{32.35} & \textbf{2.00} \\
    \bottomrule
    \end{tabular}
    \end{adjustbox}
    \caption{The main results of different LLMs on \texttt{AppBench}. Bold highlights the best score among all models, and underline underscores the best score under the same model scale}
    \label{tab:main_exp}
\end{table*}

\subsection{Evaluation Metrics}
In order to evaluate the LLMs' capabilities of selecting proper APPs, choosing APIs, and fulfilling all arguments to execute the API based on the users' instruction, we carefully design two F1 scores for APP and API, and one overall success rate considering the complexity of the task. We also provide the results of EM metrics in the Appendix.

% We expect to achieve more accurate evaluation results.
% To alleviate the bias of the hit rate, we use the following three F1-based matrics.

\paragraph{F1 of App.} We first get the precision \(P_{app}\) as the number of correctly predicted APPs divided by the total number of APPs predicted by the model:

\begin{equation}
\begin{aligned}
    P_{app} &= \frac{app\_hit\_num}{app\_pred\_num} 
\end{aligned}
\end{equation}

and recall \(R_{app}\) as the number of correctly predicted APPs divided by the total number of APPs that are in the ground truth as follows. 
\begin{equation}
\begin{aligned}
    R_{app} &= \frac{app\_hit\_num}{app\_ground\_truth\_num}
\end{aligned}
\end{equation}
\noindent The F1 of App score is  2PR / (P+R), as usual.
\paragraph{F1 of API.} 
Similarly, the metrics of API predictions can be evaluated using \(F1_{api}\). Note that we only consider the name of the API here to determine LLM whether or not to choose the right API, and the performance of arguments of APIs is evaluated in the next metric.

\paragraph{Success Rate (Succ):}
This metric evaluates whether the LLMs can fully execute the user's instruction by correctly identifying all required APPs, APIs, and arguments. It is defined as the proportion of instances where all elements—APP, API, and arguments—are in perfect alignment with the ground truth, considering the complex dependency relationship between different APIs across APPs, resulting in a direct measure of model capability in full instruction fulfillment. Since there may exist different output orders, we calculate this at the structure level since the execution structure is unique.

% Note that our evaluation does not encompass the validation of return arguments. 
% This exclusion is based on the fact that these arguments are generated by the APIs themselves and are inherently accurate, such as the results produced by a SQL request API. 
% Accordingly, our analysis assumes these return values to be correct by default.

\subsection{Main Results}

Table~\ref{tab:main_exp} shows the results of different LLMs for different types of user instructions on \texttt{AppBench}, respectively. Several conclusions can be drawn from the results \footnote{The conclusions are consistent with EM results at Table~\ref{tab:main_exp_em}.}.

\paragraph{\textit{Overall, GPT-4o achieves the best overall performance, while LLaMA3-70B sometimes outperforms GPT-3.5, mostly in scenarios only involving single APP}.} In general, other models significantly lag behind GPT-4o in all types of instructions, and only QWen1.5-72B or LLaMA3-70B achieves better or competitive performance compared with GPT-4o. Despite significant advancements in LLMs, the existing models still fall short in addressing the complexities of planning cases such as multiple APPs and multiple APIs. One fact is that all LLMs only get less than 3\% Succ in MM situations.

\paragraph{\textit{As the size of the model increases, the performance can get further improved regardless of the type of instructions and the improvement becomes less significant with multiple APPs.}} As evidenced by LLaMA3 and QWen1.5 series models, we can find that large models mostly lead to better performance. However, when the instruction requires coordination between multiple APPs, most models show a significant drop in performance and some models even get 0 at Succ, such as QWen1.5-7B and 14B. Moreover, the $F1_{app}$ can get around 10\% improvement in a single APP while only less than 5\% in LLaMA3 series models. 

\paragraph{\textit{The complexity of planning highly impacts the performance of these models.}} From the varying scores of different LLMs across different scenarios, a trend in performance emerges: the observed order of performance is approximately: MM < MS < SM < SS. This trend exists in most LLMs such as GPT-4o, QWen1.5-14B, LLaMA3-8B, and LLaMA3-70B. The slight difference between SM and MS can be attributed to different percentages of specific data examples such as the number of APPs and APIs. This kind of trend also aligns well with our intuition that the MM scenario is the most complicated, followed by MS and SM, and SS is the simplest.

\section{Analysis}
In this section, we conduct a comprehensive analysis, aiming to answer three research questions. \textbf{RQ1}: \textit{How do the parallel and sequential dependencies influence the model performance?} (Sec~\ref{sec_structure}) \textbf{RQ2}: \textit{Is it necessary to identify APP first to reduce the context window?} (Sec~\ref{sec_different_prompt}) and \textbf{RQ3:} \textit{What is the major bottleneck of current LLMs} (Sec~\ref{sec_error_ana}), \textit{and can fine-tuning or in-context learning alleviate it?} (Sec~\ref{sec_finetune}, ~\ref{sec_in_context}).

\subsection{The Effects of Dependency Structures}
\label{sec_structure}

We classify the dependency structures among APIs as twofold: parallel execution and sequential execution. For each data sample, we measure the {parallel} execution scale by the number of connected components of APIs and use the average size of these API-connected components as the {sequential} execution scale.
The data sample with a sequential scale of $1$ means no sequential dependencies among APIs. All of the APIs can be finished in a parallel way.
Then, we classify the data samples of \texttt{AppBench} based on the above criteria and discard the categories with less than 10 samples.

% First, we investigate the relationship between the number of APPs used by a user instruction and the dependency length of the used APIs. 

% We categorize data samples of AppBench based on the unique APPs numbers and the average size of these API-connected components, which acts as a measure of \textbf{sequential} execution scale.

% By fixing the number of APPs, we vary the {sequential} execution scale to explore how longer \textbf{sequential} execution paths affect performance. 
% Conversely, by fixing the API dependency length, we vary the number of APPs to examine the impact on performance when the number of APPs that need to be executed in \textbf{parallel} increases under similar execution path lengths.

\begin{figure}[t]
    \centering
    \includegraphics[trim={0cm 0cm 0cm 1.5cm}, clip, width=0.4\textwidth]{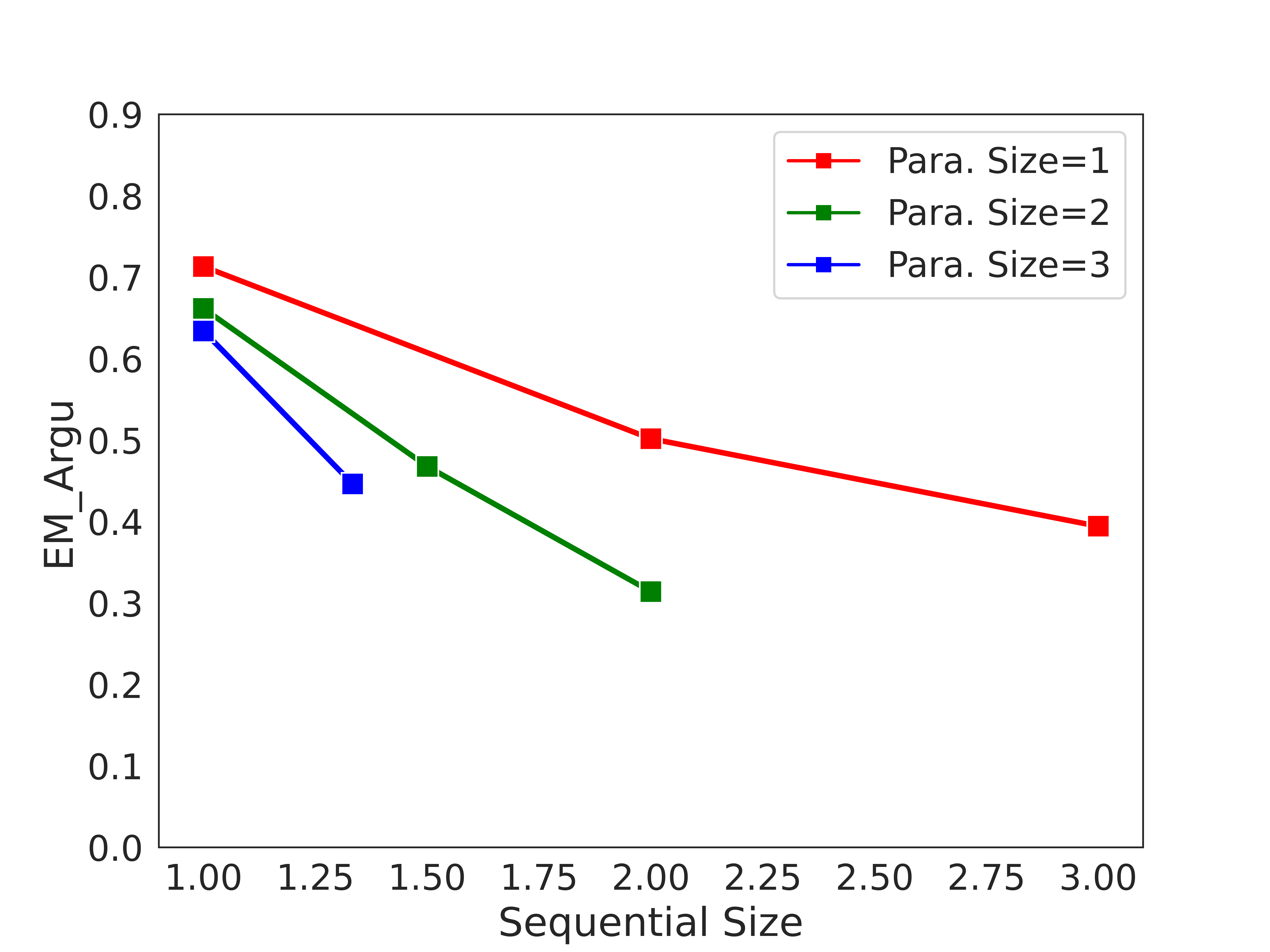}
    \caption{The relationship between GPT-4o's performance with parallel and sequential scaling. Both parallel and sequential scaling cause challenges for model performance.}
    \label{fig:comp_seq}
\end{figure}

We illustrate the Exact Match (EM) of Arguments of GPT-4o in Figure~\ref{fig:comp_seq} since arguments are directly related to the dependency relationship. First of all, when the parallel scale is fixed, an increased sequential scale becomes more challenging for GPT-4o, and vice versa. Secondly, GPT-4o appears to struggle more with sequential-complex data than parallel-complex samples. The gap between different para. size (i.e., when seq. size is fixed) is much smaller than the gap between different seq. size (i.e., when para. size is fixed).

% disregard the boundaries between different APPs and directly consider the dependency relationships among the APIs required to complete the entire user instruction, which is a clearer structure screening.
% We measure the \textbf{parallel} execution scale by the number of connected components of APIs for each data sample and also use the average size of these API-connected components as the \textbf{sequential} execution scale.
% We explore how the model performance is affected when we fix either the parallel or sequential scale and vary the other. 

\begin{figure}[t]
    \centering
    \includegraphics[trim={0.2cm 0.2cm 0cm 0cm}, clip, width=0.42\textwidth]{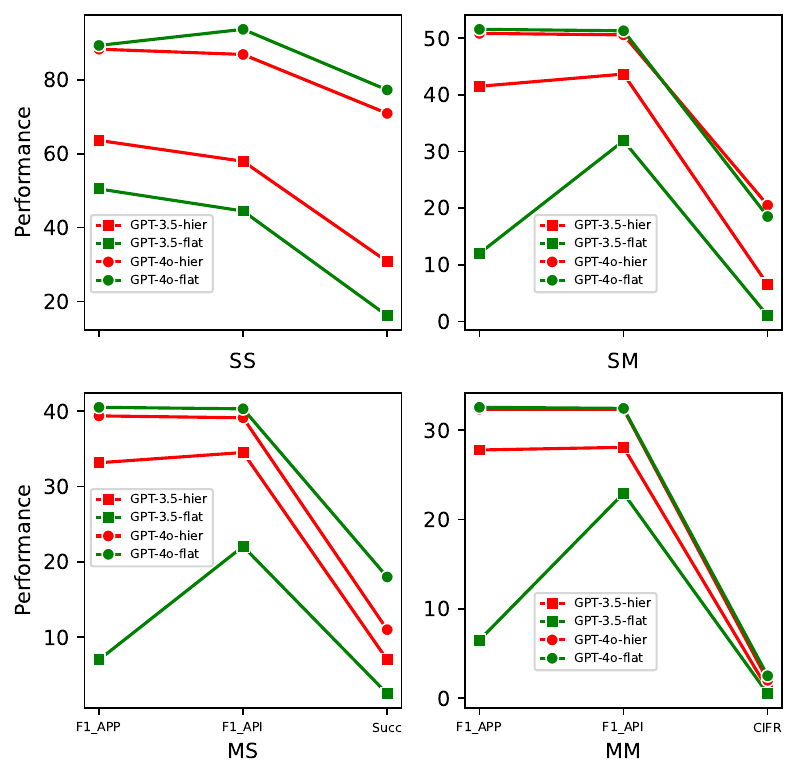}
    \captionsetup{font=10pt}
    \caption{The performance gap between hierarchical and flat prompting on GPT-3.5 and GPT-4o.}
    \label{hier_flat_comp}
\end{figure}

\subsection{The Effects of Different Prompting}
\label{sec_different_prompt}

In the main experiments, we initially required LLMs to select candidate APPs based on user input and the APP's descriptions, and then generate API calls, resulting in \textit{hierarchical} prompting. Recently, many studies have expanded the context of LLMs to 200K or more \cite{huang2024advancing}. 
Many of these works proposed LLMs with a context window that is sufficient to accommodate all the descriptions of APPs and APIs at once (\textit{flat} prompting). 
Therefore, this section explores how the model would perform if we directly provided all apps and APIs to the model. We test GPT-3.5 and GPT-4o and compare the results in Figure~\ref{hier_flat_comp}.

We can observe that flat prompting has impacted the performance of the GPT-3.5, with obvious declines in metrics such as $F1_{app}$ scores across data types. We attribute this to the introduction of a large amount of irrelevant information, which affects the model's understanding and extraction of useful APPs and APIs. Surprisingly, the GPT-4o model achieved better performance using flat prompting. We believe this is due to the GPT-4o's more powerful long-context understanding capabilities, which allow it to accurately identify the required APP and API. Moreover, the absence of the error propagation effect that occurs during the first APP selection step of hierarchical prompting, has led to a clear improvement in performance. However, flat prompting requires a strong contextual capability that few models possess, and it necessitates the input of a large number of irrelevant tokens, which incurs additional computational power consumption.

\begin{table}[!t]
\setlength{\belowcaptionskip}{0pt}
    \centering
    \begin{adjustbox}{max width=0.4 \textwidth}
    \begin{tabular}{l| cc|ccc}
    \toprule
    \multirow{2}{*}{\textbf{Category}} &  \multicolumn{2}{c}{\textbf{Keys}}  & \multicolumn{3}{|c}{\textbf{Values}} \\    
    % dynamic / multi-turn?
    \cline{2-6} & I & D & I & D & T/S \\
    \hline
    \textbf{SS} & 6.1 & - & 6.6 & - & 42.1/26.3 \\
    \textbf{SM} & 5.5 & 8.0 & 2.5 & 75.5 & 27.1/15.2 \\
    \textbf{MS} & 6.0 & 1.0 & 6.0 & 30.0 & 45.1/8.8 \\
    \textbf{MM} & 19.0 & 15.0 & 6.0 & 82.0 & 36.8/24.6 \\
    \bottomrule
    \end{tabular}
    \end{adjustbox}
    \caption{Error analysis of GPT-4o on \texttt{AppBench}. I and D stand for independent and dependent variables or values, respectively, between multiple APIs. T/S refers to time-related or space-related values, such as start date and location.}
    \label{tab:error_analys}
\end{table}

% \begin{table}[!t]
% %\setlength{\abovecaptionskip}{5pt}   
% \setlength{\belowcaptionskip}{0pt}
%     \centering
%     \begin{adjustbox}{max width=0.48 \textwidth}
%     \begin{tabular}{l| ccc|ccc}
%     \toprule
%     \multirow{2}{*}{\textbf{Type}} & \multicolumn{3}{c}{\textbf{Indep. Argus.}}  & \multicolumn{3}{|c}{\textbf{Dep. Argus}} \\    
%     % dynamic / multi-turn?
%     \cline{2-7} & $F1_{iargu}$ & $R_k$ &$R_v$ & $F1_{dargu}$ & $R_k$ & $R_v$ \\
%     \hline
%     SS & 99.51 & 4.75 & 10.76 & - & - & - \\
%      SM &76.70 & 7.36 & 52.96 &  68.76 & 1.66 & 26.78 \\
    
%     MS &73.53 & 2.01 & 7.49 & 3.91 & 0.46 &14.68\\
%     MM & 73.43 & 9.64 & 33.99 & 63.85 & 2.55 & 20.71\\
    
%     \bottomrule
%     \end{tabular}
%     \end{adjustbox}
%     \caption{Error analysis of GPT-4o predictions.}
%     \label{tab:error_analys}
% \end{table}

\begin{table*}[t]
\setlength{\belowcaptionskip}{0pt}
    \centering
    \begin{adjustbox}{max width=0.9 \textwidth}
    \begin{tabular}{l| ccc|ccc|ccc|ccc}
    \toprule
    \multirow{2}{*}{\textbf{Settings}} & \multicolumn{3}{c}{\textbf{SS}}  & \multicolumn{3}{|c}{\textbf{SM}} & \multicolumn{3}{c}{\textbf{MS}}  & \multicolumn{3}{|c}{\textbf{MM}} \\    
    % dynamic / multi-turn?
    \cline{2-13} & $F1_{app}$ & $F1_{api}$ & Succ & $F1_{app}$ & $F1_{api}$ & Succ & $F1_{app}$ & $F1_{api}$ & Succ & $F1_{app}$ & $F1_{api}$ & Succ \\
    \hline
    GPT-4o & 88.31 & 86.87 & 70.92 & 50.83 & 50.57 & 20.50 & 39.39 & 39.14 & 11.00 & 32.36 & 32.35 & 2.00 \\
    \hdashline
    \textit{3-shot} & 93.73 & 90.73 & 81.63 & 51.16 & 50.90 & 13.50 & 40.12 & 39.92 & 12.50 & 32.72 & 32.73 & 2.50 \\
    \textit{4-shot} & 93.23 & 89.72 & 79.59 & 50.96 & 50.70 & 14.00 & 40.29 & 40.29 & 10.50 & 32.44 & 32.44 & 3.00\\
    \textit{5-shot} & 93.70 & 91.18 & 79.59 & 50.32 & 50.06 & 14.00 & 40.33 & 40.12 & 12.50 & 32.36 & 32.36 & 2.50 \\
    \bottomrule
    \end{tabular}
    \end{adjustbox}
    \caption{In-context learning results of GPT-4o on \texttt{AppBench}.}
    \label{tab:in_context}
\end{table*}

\subsection{Error Analysis}
\label{sec_error_ana}

We further conduct error analysis at the argument level since it is directly related to different relationships between multiple APIs, to identify potential bottlenecks of the current best model: GPT-4o. Specifically, there are two main categories of errors to consider: \textit{1) key error.} It occurs when the model predicts fewer keys than expected to successfully execute the API call, and it can be further divided into two types: \textbf{I}ndependent: The missing or incorrect keys are from the independent variables or arguments and \textbf{D}ependent: The missing or incorrect keys are from the dependent variables or arguments; and \textit{2) value error}. it occurs when the model predicts values that do not match the ground truth values, given the name of the key.
Value errors can also be divided into I and D types.

Table~\ref{tab:error_analys} presents the percentage of error cases over the number of total arguments in each category while T/S is the percentage over all error arguments in each category. It is found that as complexity increases, errors also increase. The lower D-key error and D-value error in MS can be attributed to a smaller percentage of dependency cases in this category. Out of all types of errors, the D-value error appears to be the biggest bottleneck or challenge for the LLM. Further analysis reveals that the value errors are particularly prevalent for time and space-related keys. For example, the language models may struggle to accurately recognize or reason about date/time expressions used in the user's input, such as "next Monday".

\subsection{The Effects of Fine-tuning}
\label{sec_finetune}

\begin{figure}[t]
    \centering
    \includegraphics[trim={0cm 0cm 0cm 0cm}, clip, width=0.48\textwidth]{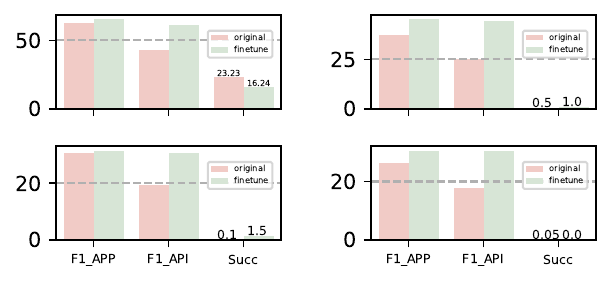}
    \captionsetup{font=10pt}
    \caption{The performance gap between original LLaMA3-8B and finetuned one.}
    \label{comp_origin_finetune}
\end{figure}

We additionally collected around 1,000 samples for each category from the training dataset of SGD, resulting in 4,000 samples total. We then used this mixed dataset to fine-tune the LLaMA3-8B model. Figure~\ref{comp_origin_finetune} shows the final results. Further fine-tuning on in-domain data did bring some improvement in the F1 score of the APP and API, but can not boost the performance for the Succ. Upon closer inspection, we found that the major reason for the lower performance on Succ was due to issues with recognizing or matching the keys and values in the input arguments. The model sometimes failed to recognize all the necessary input keys and values, or mistakenly used keys from other APIs. This appears to be strongly related to the complexity of the task. Factors like the dependency relationships between multiple APIs, as well as the lengthy API descriptions, made it challenging for the LMs to fully capture the necessary patterns and logic. 

% The complexity of the task seems to be the primary limiting factor in achieving higher performance, call for more exploration on effective finetuning methods.

\subsection{The Effects of In-context Learning}
\label{sec_in_context}

Table~\ref{tab:in_context} shows the in-context performance of GPT-4o when using different shots at the demonstrations. Specifically, we randomly sample (instruction, outputs) from the same training set created during fine-tuning according to the used APP in the current instruction. For example, if the used APP in current instruction is Hotel, we sample the first 3 appeared samples with the same APP in the training set to form 3-shot demonstrations, aiming to save the space of additional API descriptions and make the agent familiar the utilization of current API. From the table, we find that in-context learning shows some improvement in simpler cases, such as SS ($\approx$ 10 points increase on Succ). However, the performance does not improve further as situations become more complex and even decreases in scenarios like SM or MS, highlighting the challenges of complex planning. The worst performance in SM may be strongly related to our sampling strategy, as we only consider the APP level rather than different APIs within the same APP. More effective in-context learning for complex planning is desired and warrants further exploration and attention.

% \noindent \textbf{Privacy Protection} It is necessary to consider the setting where details about APIs inside each APP are protected by the developers and not accessible by the meta agent, such as the API name and related definitions of arguments. 
% In this way, the APP is a black box with only the required input, and the returned output is known by the meta agent. 

\section{Conclusion}
In this paper, we introduce a new benchmark, \texttt{AppBench}, addressing the challenge of complex user instructions that require the involvement of multiple APIs. These scenarios demand advanced planning capabilities from LLMs to effectively handle graph structures and ensure permission isolation in practical applications. We left the self-evolving or more effective fine-tuning framework in our future work.

\section{Acknowledgement}

Thanks for the insightful comments and feedback from the reviewers. This work was supported by the National Key R\&D Program of China (No. 2023YFF0725600) and the National Natural Science Foundation of China (No. 62406015). This research work also is partially supported by CUHK direct grant (No. 4055209) and CUHK Knowledge Transfer Project Fund No. (KPF23GWP20).

\section{Limitations}
% Though multiple valid solutions might exist for a given instruction in practice, this study limits its focus to a single anticipated solution due to the ToD dataset only including one solution per dialogue, which constrains the AppBench dataset.

We acknowledge the following limitations in terms of the evaluations and benchmarks.

\paragraph{Evaluations.} We do not consider the existing agent framework since we mainly focus on the base capabilities of various LLMs on this new benchmark. We anticipate that introducing additional reflection or a carefully designed agent framework may further boost the performance of original LLMs.

\paragraph{Benchmarks.} We mainly take advantage of existing task-oriented datasets to build our benchmark, which brings two limitations: 1) We main focus on text-based natural language interactions while the API also works at different modalities. We leave this in future work, and 2) We do not consider APPs with overlap or similar functions, but they exist in practice such as different platforms to buy tickets. We argue that these apps can often be distinguished through minor modifications to the app names and APIs. The specific choice of which app a user selects ultimately comes down to individual user preferences, which is outside the scope of this paper.

% 1) existing LLMs may unintentionally use this dataset during pre-training or fine-tuning due to data contamination issues \citep{schaeffer2023pretraining, zhang2024careful}, leading to illusory higher performance on specific LLMs; 2) We mainly focus on single-turn interaction, following \cite{qin2023toolllm, huang2024planning}, while the multi-turn interaction also plays key role in daily life;

\section{Ethical Considerations}

In conducting our research, we have thoroughly reviewed and ensured compliance with ethical standards. Our study utilizes existing datasets, which have been publicly available and previously vetted for ethical use. These datasets have been carefully selected to avoid any form of offensive or biased content. Therefore, we consider that our research does not present any ethical issues. The data used is ethically sourced, the analysis is unbiased, and all procedures align with established ethical guidelines.

\bibliography{custom}
\bibliographystyle{acl_natbib}
\clearpage

\appendix

\section{Data Collection}
\label{appendix:data_collection}

\subsection{Prompt Details}

\begin{table}[ht]
\small
    \centering
    \colorbox{blue!8}{
    \begin{tabular}{@{}p{7.2cm}}
    Your task is to generate a complex instruction in one sentence which exactly reflect what user want to do during the dialogue with the dialogue system as follows.  \\\\

    Please give all specific values of user requirements in user aware arguments \{\texttt{user\_aware\_arguments}\}.  \\\\
    
    You should not know any values of other arguments specified by the system side.
    \end{tabular}
    }
    \caption{The prompt used to prompt LLM to generate the summarized instruction}
    \label{table:data_generation_prompt}
\end{table}

\begin{table}[ht]
\small
    \centering
    \colorbox{blue!8}{
    \begin{tabular}{@{}p{7.2cm}}
    Please evaluate the given instruction based on the following criteria: \\\\

    Fluency:
    
    -- Evaluate the prompt's clarity, coherence, and ease of understanding. \\
    -- Consider factors such as the organization, language flow, and presentation of the prompt. \\\\
    
    Diversity:
    
    -- Evaluate the range of topics, perspectives, and related APP and APIs covered by the prompt. \\\\
    
    Please only output the overall score considering both of fluency and diversity. The overall score should be a value between 1 and 10, with 10 representing the best. \\
    \end{tabular}
    }
    \caption{Prompts to evaluate the quality of generated instructions.}
    \label{tab:quality_score}
\end{table}

\subsection{Data Statistics}
\label{appendix_data_stat}

\begin{table*}[h]
\centering
    \begin{adjustbox}{max width=0.9 \textwidth}
    \begin{tabular}{l | cccc}
    \hline
    \textbf{App} & \textbf{APIs} \\
    \hline
    Rents & \textit{getcarsavailable}, \textit{reservecar}, \textit{getride} \\
    Hotels & \textit{searchhouse}, \textit{bookhouse} \\
    Services & \textit{book\_stylist\_appointment}, \textit{find\_stylist\_provider}, \textit{book\_therapist\_appointment}, \textit{find\_therapist\_appointment} \\
    Restaurant & \textit{reserverestaurant}, \textit{findrestaurants} \\
    Movies & \textit{buymovietickets}, \textit{findmovies}, \textit{gettimesformovie}, \textit{reviewmovies} \\
    Trains &  \textit{gettraintickets}, \textit{findtrains} \\
    Events & \textit{findevents}, \textit{buyeventtickets} \\
    Travel & \textit{findattractions} \\
    Buses &  \textit{findbus}, \textit{buybusticket} \\
    Flights & \textit{searchonewayflight}, \textit{searchroundtripflights} \\
    Payment & \textit{requestpayment}, \textit{makepayment} \\
    Music & \textit{playmedia}, \textit{lookupmusic} \\
    Weather & \textit{getweather} \\
    \hline
    \end{tabular}
    \end{adjustbox}
\caption{List of All Apps and their corresponding APIs in the \texttt{MetaBench}.}
\label{tab:apps_apis}
\end{table*}

\paragraph{Definition of Parallel and Sequential}
\label{appendix_para_seq}

In this section, we delve into the execution logical structure inherent in each data sample to have a better understanding of task complexity.
We conceptualize the APIs used within a single data instance as nodes within a directed graph and the dependency among them as the directed edges.
Consequently, we analyze the interrelations among all APIs within the data sample and construct a corresponding graph for each of them.
It is important to notice that not all APIs within a sample are interdependent,  and some may operate independently. 
As a result, the APIs within the same data sample generally form several distinct \textbf{components}. 
We treat each component as a unit to perform topological sorting. The execution process of each unit can be parallel, while the procedure within the component is sequential.

As shown in Figure~\ref{data_examples}, the illustrated MM sample needs to leverage 2 APIs from APP-1 and 3 APIs from APP-2 to fulfill the user instruction.
Though APIs of APP-1 are interdependent, they do not need results from APIs of APP-2. Hence the formed graph of this sample has 2 components, with a size of 2 and 1.
These two components can be fulfilled simultaneously, but the results from APIs of APP-1 or APP-2 need to be executed one by one.

Above all, to quantify the complexity inherent in these interactions, we compute both the average and maximum \textbf{sizes} of these components as sequential scale(\textit{Max. and Avg. Seq.} in Table~\ref{tab:data_stat}), which reflect the complexity of sequential dependency among the APIs. 
Additionally, we measure both the average and the maximum \textbf{number} of components within each sample (\textit{Max. and Avg. \# Para.} as the parallel scale,
% in Table~\ref{tab:data_stat}), 
providing insight into the level of parallelism among the APIs or components. 

As shown in Table~\ref{tab:data_stat}, the instances of SS and SM are relatively simple.
Since the samples of MS have the most complex sequential scales. 
The samples of MM are the most complex since their parallel and sequential scales are relatively larger than the others.

\section{Experimental Details}

\begin{table}[ht]
\small
    \centering
    \colorbox{orange!8}{
    \begin{tabular}{@{}p{7.2cm}}
    Your task is to determine the required App list according the description of each App and user requirements. \\\\
    
    Here is the information about all accessible Apps: \{\texttt{app\_desc}\} \\\\
        
     Make your response short and concise. Your \textbf{ONLY} need to return needed app names and your output MUST follow this format: [app1, app2, ...] \\\\

     User Instruction: \{\texttt{user\_instruction}\} 
    \end{tabular}
    }
    \caption{Prompts to select APP first.}
    \label{tab:app_selection_prompt}
\end{table}

\begin{table}[ht]
\small
    \centering
    \colorbox{orange!8}{
    \begin{tabular}{@{}p{7.2cm}}
    Your task is to generate App name and corresponding API calls to complete the user requirements according to given descriptions of all Apps and APIs.  \\\\
            
    Here is the information about all accessible Apps and corresponding APIs. \{\texttt{app\_api\_list}\} \\\\

    Your output should follow the format as follows: 
    
    app1: [returned\_argument1, returned\_argument2, ... = app1\_api1(\#argument1=value1, \#argument2=value2, ...)]
    app1: [returned\_argument1, returned\_argument2, ... = app1\_api2(\#argument1=value1, \#argument2=value2, ...)]
    app2: [returned\_argument1, returned\_argument2, ... = app2\_api1(\#argument1=value1, \#argument2=value2, ...)] \\\\
    
    Here are explanations:

    1. API Naming Convention 
    
    -- The API call format is [returned\_argument1, returned\_argument2, ... = app1\_api1(\#argument1=value1, \#argument2=value2, ...)]. \\
    -- app1 signifies the name of app1, and app1\_api1 signifies the name of api1 in the app1. \\\\

    2. Arguments
    
    -- argument1 is the first input arguments for the corresponding api, and so on. \\
    -- returned\_argument1 is the first output arguments from the corresponding api, and so on. \\
    -- Input arguments include both required and optional arguments as described in the corresponding API description of App. \\
    -- The order and names of input and returned arguments must exactly match the given description. \\\\

    3. Values of Input Arguments
    
    -- If specified by the user, replace the placeholder with the actual value. \\
    -- If not specified by the user, omit the optional arguments from the API call. \\
    -- If an argument value is dependent on another API's output, use the name of the returned argument as the value. \\
    -- There are no default values for any arguments. All required arguments must be provided by the user or through dependencies on other APIs' outputs. \\
    -- You should be careful about the date value, you need to infer it based on current date "2019-03-01". \\\\

    4. Order of Execution:
    
    -- Execute APIs in a sequence that respects their dependencies. For example, if api2 requires an output from api1, ensure api1 is executed before api2. \\
    -- Handle cases where multiple APIs' outputs are required for a single API's input by waiting for all dependent APIs to execute before calling the dependent API. \\\\

    Example:
    
    If api2 in app1 depends on the output of api1 in app1 and an optional argument is not provided by the user:
    
    app1: [output1 = app1\_api1(\#argument1=value1)] \\
    app1: [output2 = app1\_api2(\#argument2=output1)] \\\\

    If api3 in app2 requires outputs from both api1 in app1 and api2 in app1:
    
    app1: [output1 = app1\_api1(\#argument1=value1)] \\
    app1: [output2 = app1\_api2(\#argument2=output1)] \\
    app2: [output3 = app2\_api3(\#argument3=output1, \#argument4=output2)] \\\\

     User Instruction: \{\texttt{user\_instruction}\} \\
    \end{tabular}
    }
    \caption{Prompts to generate the final planning path to fulfill the user instruction.}
    \label{tab:planning_path_output}
\end{table}

\begin{table*}[!t]
\setlength{\belowcaptionskip}{0pt}
    \centering
    \begin{adjustbox}{max width=0.98 \textwidth}
    \begin{tabular}{l| cc|cc|cc|cc}
    \toprule
    \multirow{2}{*}{\textbf{Models}} & \multicolumn{2}{c}{\textbf{SS}}  & \multicolumn{2}{|c}{\textbf{SM}} & \multicolumn{2}{|c}{\textbf{MS}}  & \multicolumn{2}{|c}{\textbf{MM}} \\    
    % dynamic / multi-turn?
    \cline{2-9} & APP & API & APP & API & APP & API & APP & API \\
    \hline
    Mistral-7B & 27.27 & 14.14 & 19.50 & 4.50 & 1.50 & 1.00 & 2.00 & 0.00 \\
    Vicuna-13B & 31.82 & 21.21 & 7.00 & 3.00 & 1.50 & 0.50 & 0.00 & 0.00 \\
    \hdashline
    
    LLaMA3-8B & 47.98 & 47.47 & 19.00 & 17.50 & 12.50 & 9.50 & 4.50 & 5.50 \\
    LLaMA3-70B & 60.94 & \underline{58.33} & \underline{51.00} & \underline{49.00} & 12.00 & 6.50 & 16.00 & 8.50 \\
    
    \hdashline
    QWen1.5-7B & 28.28 & 12.63 & 11.50 & 4.00 & 2.50 & 0.50 & 4.00 & 1.50 \\
    QWen1.5-14B & 56.57 & 41.92 & 10.50 & 10.00 & 5.60 & 4.00 & 1.50 & 1.50  \\
    QWen1.5-72B & \underline{71.88} & 32.29 & 38.50 & 9.50 & 2.47 & 1.85 & 4.50 & 3.50 \\
    
    \hdashline
    
    GPT-3.5 & 44.44 & 52.02 & 30.50  & 31.00  & \underline{31.00}  & \underline{19.00}  & \underline{18.50} & \underline{19.50}   \\
    GPT-4o & \textbf{79.59} & \textbf{78.06} & \textbf{55.50} & \textbf{51.50} & \textbf{35.50} & \textbf{26.50}  & \textbf{29.50 }& \textbf{24.00} \\
    \bottomrule
    \end{tabular}
    \end{adjustbox}
    \caption{The Exact Match (EM) results of different LLMs on \texttt{MetaBench}. Bold highlights the best score among all models, and underline underscores the best score under the same model scale}
    \label{tab:main_exp_em}
\end{table*}

\end{document}